\documentclass{Interspeech2024}

\interspeechcameraready

\usepackage{graphicx}
\usepackage{multicol} 
\usepackage{multirow}
\usepackage{amsmath}
\usepackage{subcaption}

\title{Emotional Cues Extraction and Fusion for Multi-modal Emotion Prediction and Recognition in Conversation} 

\name[affiliation={\dagger}]{Haoxiang}{Shi}
\name[affiliation={\dagger}]{Ziqi}{Liang}
\name[affiliation={\ast}]{Jun}{Yu}

\address{University of Science and Technology of China, Hefei, China\thanks{$\dagger$ Equal Contribution}\thanks{$\ast$ Corresponding author: Jun Yu}}

\email{harryjun@ustc.edu.cn}


\begin{document}

\maketitle

\begin{abstract}
    
    Emotion Prediction in Conversation (EPC) aims to forecast the emotions of forthcoming utterances by utilizing preceding dialogues. Previous EPC approaches relied on simple context modeling for emotion extraction, overlooking fine-grained emotion cues at the word level. Additionally, prior works failed to account for the intrinsic differences between modalities, resulting in redundant information. To overcome these limitations, we propose an emotional cues extraction and fusion network, which consists of two stages: a modality-specific learning stage that utilizes word-level labels and prosody learning to construct emotion embedding spaces for each modality, and a two-step fusion stage for integrating multi-modal features. Moreover, the emotion features extracted by our model are also applicable to the Emotion Recognition in Conversation (ERC) task. Experimental results validate the efficacy of the proposed method, demonstrating superior performance on both IEMOCAP and MELD datasets.
\end{abstract}

\section{Introduction}
Emotions play an important role in human communication\cite{mim,cau}. Proper understanding and application help us communicate better and establish connections. In human-machine dialogue, it is also crucial to understand the user's emotional changes. Emotion prediction can enable intelligent agents to achieve more friendly and considerate interactions, improving user experience and human-machine interaction efficiency.

Emotion prediction in conversation (EPC) involves forecasting the speaker's future emotional state using preceding contextual cues\cite{2}. Previous research have focused on modeling historical context information in sessions to improve the accuracy of EPC task\cite{3,Shi20,5}. Shi et al.\cite{Shi20} performed subsequent emotion prediction by correlating the contextual speaking information of bilateral speakers instead of a single speaker. In their subsequent work\cite{Shi23}, Shi et al. further considered the importance of multi-turn dialogues and modeled them, improving EPC task accuracy. However, word-level associations often constitute the core of conversational dynamics. Thus, to facilitate comprehensive comprehension and subsequent predictions, modeling contextual word-level relationships is imperative.

In addition, regarding the utilization of multi-modal information, previous EPC approaches applied identical modeling techniques to both text and audio modalities without considering their distinct characteristics. The text modality typically encapsulates clear content and character information, whereas the audio modality includes not only the spoken content but also valuable speaker-specific and prosodic cues\cite{aia}. Therefore, it is crucial to extract and enhance distinctive features within each modality prior to their fusion. Furthermore, during modal fusion, prior models\cite{concat1,concat2} often directly concatenated modalities, thereby overlooking the potential interactions and complementary information between them.

To address these challenges, we propose a multi-step fusion model based on enhanced intra-modal emotional information to predict the emotions of multi-party conversations. Specifically, the model consists of two stages. In the first stage, we perform intra-modal emotional cues perception and enhancement. For text, we introduce a knowledge-based word relation tagging module, which helps the model pay attention to word-level emotional features and improve emotional perception capabilities by constructing a word-level importance matrix within the conversation. As for speech, we design a prosody enhancement module aimed at detecting prosodic changes to better predict emotional trends in conversations. In the second stage, we employ a two-step fusion method to integrate multi-modal features. Initially, we fuse the intra-modal features obtained in the first stage to generate the preliminary fusion representation. Subsequently, the representation fused again with the mel-spectrogram extracted from the audio waveform to incorporate spectral domain emotional features and generate the final fused emotional representation. Additionally, we conducted experiments on the Emotion Recognition in Conversation (ERC) task to validate the effectiveness and multi-task applicability of our model's emotional feature extraction and fusion. The main contribution of this work can be summarized as follows: 
\begin{itemize}
\item We propose a novel method to complete the EPC task, which perceives the intra-modal features through word-level relation tagging and prosody enhancement modules. 
\item We introduce a novel multi-modal fusion method that leverages the spectral domain features of audio for two-step multi-modal information fusion.
\item We conduct experiments on both EPC and ERC tasks, experimental results demonstrate that our model achieves leading performance in two tasks.
\end{itemize}

\begin{figure*}[ht]
\begin{minipage}[b]{1.0\linewidth}
  \centering
  \centerline{\includegraphics[width=16cm]{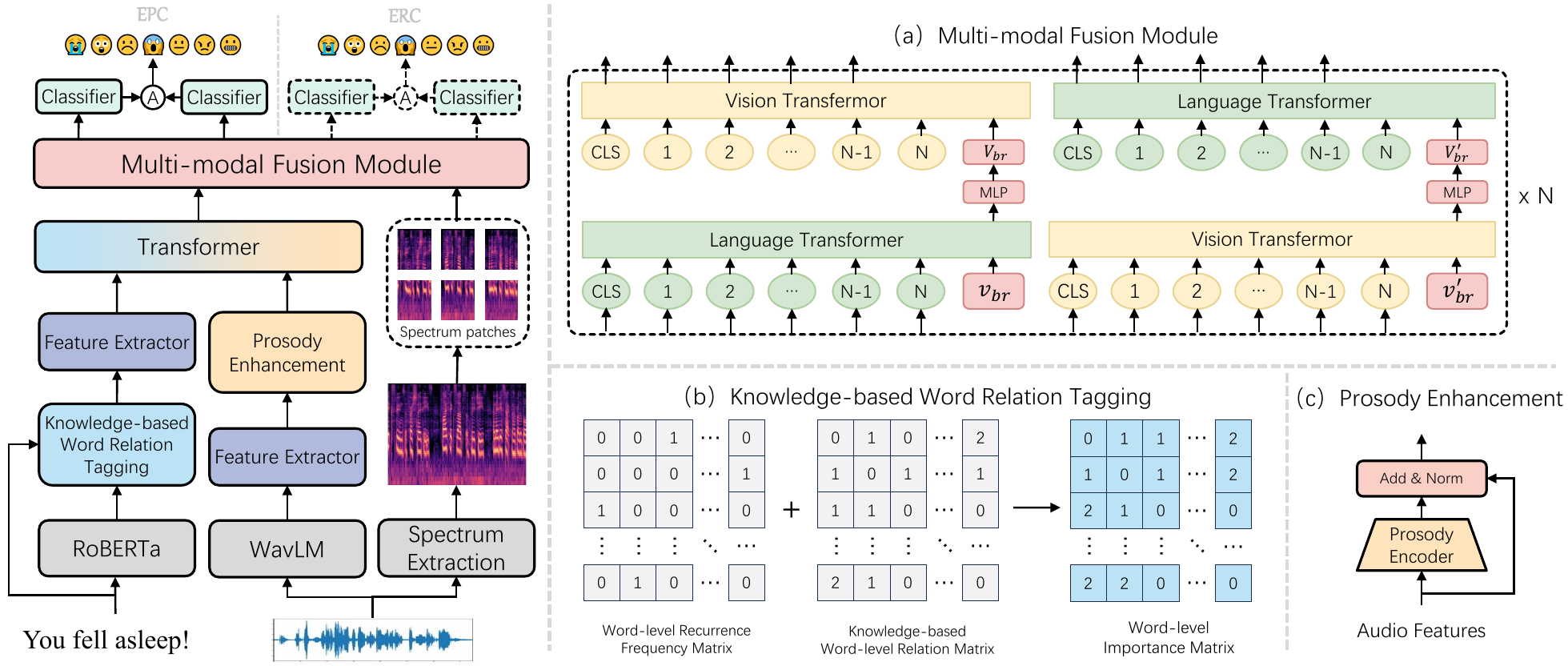}}
  \caption{The left side is the main framework of our model. (a) multi-modal fusion module, the green part represents the pre-trained language transformer layer, the yellow part represents the pre-trained vision transformer layer, the pink part is trainable, and other parts are frozen. It is worth noting that, in (b), we first initialize the the word-level relation matrix $M_{rel}$ with 0. For each word pair, if a relation exists, the importance is increased by 1, with a maximum upper limit of 3.}
  \label{arc}
\end{minipage}
\end{figure*}

\section{Methods}
\subsection{Task Definition}
\textbf{Emotion Prediction in Conversation (EPC)} In a multi-modal multi-party (or dyadic) dialogue containing text and audio $\mathbb{D} = \{(u_1, s_1), (u_2, s_2), ..., (u_N, s_N)\}$, where $(u_i,s_i)$ represents the $i^{th}$ utterance-speaker pair in the conversation, N is the number of utterances in the dialogue. EPC aims to predict the emotion category label $e_{n+1}$ of the future utterance-speaker pair $(u_{n+1}, s_{n+1})$ by given the historical dialogues $\{(u_1, s_1), (u_2, s_2), ..., (u_{n}, s_{n})\}$. \vspace{5pt}\\
\textbf{Emotion Recognition in Conversation (ERC)} In the same conversation $\mathbb{D}$, differnt from EPC, ERC task requires identifying the emotion category labels $E=\{e_1, e_2, ..., e_N\}$ of the dialogue $\{(u_1, s_1), (u_2, s_2), ..., (u_N, s_N)\}$. 

\subsection{Overview}
The architecture of our model, as depicted in Figure~\ref{arc}, is primarily composed of three parts: the intra-modal emotion perception part, the two-step multi-modal fusion part, and the final emotion classification part. Specifically, in the intra-modal emotion perception part, we design a Knowledge-based Word Relation Tagging (KWRT) module and a Prosody Enhancement (PE) module to address the differences in information between modalities. In the two-step multi-modal fusion part, we first fuse the text and audio features enhanced in the first part, then combine the initial fusion features with the mel-spectrogram extracted from the audio to generate the final multi-modal fusion representation. Finally, we obtain the prediction and recognition results by designing classifiers for the EPC and ERC tasks in the emotion classification part. 

\subsection{Knowledge-based Word Relation Tagging (KWRT)}
Given a conversation, we first transform it into a sequence based on utterances and speaker information, and input it into RoBERTa\cite{roberta} to discern the dynamics and interdependencies among speakers within the conversation. Specifically, we distinguish different speakers through different special tokens, such as $[s1]$ and $[s2]$. Then we get a conversation sequence $x$ containing speakers and corresponding utterances. Now assume that the number of tokens in $x$ is $N$, and then we use RoBERTa to encode the utterance sequence: $h_t = RoBERTa(x)$, where $h_t\in \mathbb{R}^{d \mathrm{x} N}$ and d is the output dimension of RoBERTa.

\begin{figure}[h]
\label{fig2}
\begin{minipage}[b]{1.0\linewidth}
  \centering
  \centerline{\includegraphics[width=8cm]{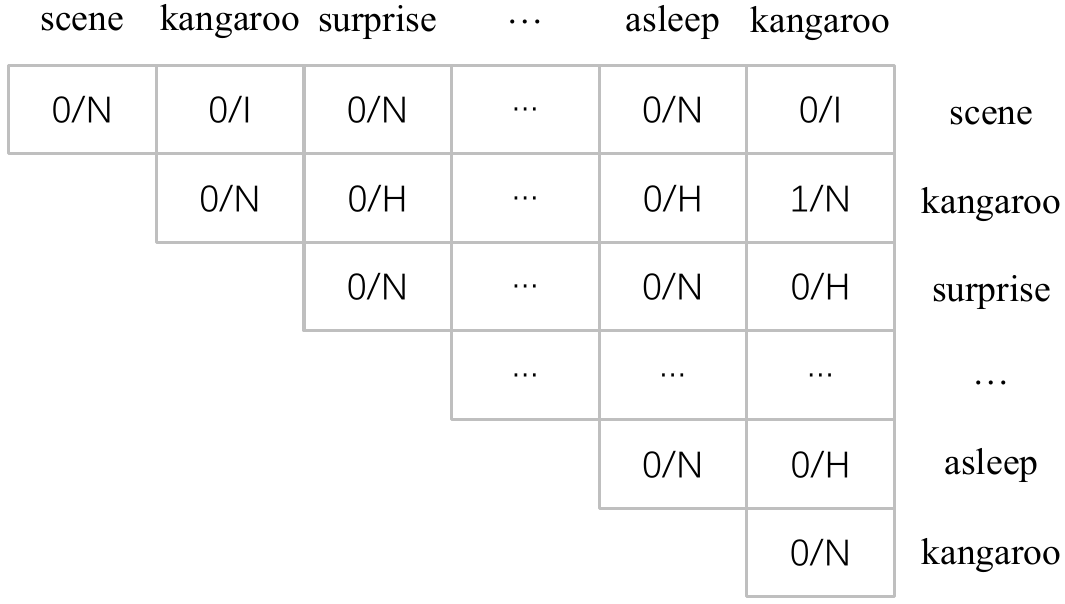}}
  \caption{A tagging example with KWRT module.}
\label{f2}
\end{minipage}
\end{figure}

Our goal is to explore word-level relationships between contextual dialogues, inspired by \cite{dd,dd1}, we design a word relation tagging method to obtain the importance level of words, thereby enabling the model to extract emotional cues. Taking the example of the 3-turn dialogues, we first concatenate the three segments of dialogue to construct a word-level matrix, as shown in Figure~\ref{f2}. We exclude function words, retaining only content words, as only the word-level relationships between content words contain emotional content information. We use two word-level tags to assess word importance comprehensively: word recurrence and word relations. For example, we use ``0/H'' to represent the relationship between the word pair ``asleep-kangaroo''. Here, `0' indicates the frequency of word recurrence. If two words are identical and located off-diagonally, they are labeled as `1', indicating that the words have appeared at least twice in a dialogue; otherwise, they are labeled as `0'. Meanwhile, `H' represents the lexical relationship between them. As emotional cues within dialogues alone cannot provide knowledge beyond the context, we use the common-sense knowledge base ConceptNet\cite{conceptnet} as an external source to tag relationships between word pairs. We select three main relations among 38 pairs, namely ``IsA,'' ``HasContext,'' and ``Causes,'' which may contain emotional cues, and we use `I', `H' and `C' to represent them, respectively. It should be noted that word pairs containing function words and those on the diagonal are marked as ``0/N''.

At this point, by assigning word-level labels, we obtain a word-level recurrence frequency matrix $M_{rec}$ and a word-level relationship matrix $M_{rel}$. Here, $K$ denotes the total number of words in the dialogue. We then superimpose the two matrices to obtain the final word-level importance matrix $M$, as shown in Figure~\ref{arc}b, where $\{M_{rec}, M_{rel}, M\}\in \mathbb{R}^{K\mathrm {x}K}$. 

Finally, we combine the word-level importance matrix $M$ with utterance-level features$h$:
\begin{equation}
\begin{aligned}
\mathcal{F}_t = h_t * Linear(Squeeze(M))
\end{aligned}
\end{equation}
where $\mathcal{F}_t\in \mathbb{R}^{d\mathrm{x}K}$ is the final utterance representation. The operation $Squeeze(M)$ compresses the dimensions of the $M$ matrix into $\mathbb{R}^{1\mathrm {x}K}$ since each row in M corresponds to the importance score vector of a word.

\subsection{Prosody Enhancement (PE)}
In previous emotion prediction models\cite{pe1,pe2}, features were often extracted from audio using the same processing methods as text and then fused with text features. However, these approaches did not account for the unique information inherent in each modality\cite{aia}. In the text modality, we consider the semantic connections between content words. However, the expression of sentiments associated with function words, such as ``what'' or ``where'' in interrogative sentences, is often conveyed through the prosody in audio. Therefore, we design a prosody enhancement module to extract and amplify emotional cues from audio features. Specifically, we first extract preliminary audio features $h_a$ using a pre-trained audio model and a feature extraction module. We then use $h_a$ to extract emotional cues via the fine-tuned prosody encoder as described in \cite{enc}. Similar to the text modality, we derive the final audio emotional feature representation by integrating the prosody information with the preliminary audio features:

\begin{equation}
\begin{aligned}
\mathcal{F}_a = h_a + LN(\mathbf{Enc}(h_a))
\end{aligned}
\end{equation}
where $\mathbf{Enc}()$ stands for prosody encoder, $LN()$ represents the Layer Normalization.

\subsection{Two-step Multi-modal Fusion Layer (TMF)}
To integrate the bimodal features of text and audio, we adopt a two-step fusion strategy. Initially, we utilize a Transformer\cite{trans} to integrate multi-modal emotional features. The audio features are treated as the query matrix, while the text modal features are considered as the key-value matrix, enabling the propagation of information between modalities, resulting in preliminary complementary fusion features, denoted as $\mathcal{F}_{t,a}$.

Besides the pitch contour, the mel-spectrogram also carries prosodic information\cite{proso}. Therefore, to further leverage the frequency domain information of audio, we design a multi-layer multi-modal fusion module (MFM) utilizing the mel-spectrogram, as depicted in Figure~\ref{arc}a. We employ pre-trained large-scale language\cite{bert} and vision\cite{vit} Transformer models to process the fusion and spectral domain information, represented as $Trans_l$ and $Trans_v$, respectively. Initially, according to \cite{pmf}, we initialize a trainable bridge vector, denoted as $\mathbf{v}_{br}$, to facilitate the multi-modal fusion process. For instance, taking the spectral domain features $\mathcal{F}_{m}$ as the initial input, we concatenate $\mathbf{v}_{br}$ with $\mathcal{F}_{m}$, and subsequently feed them into a unimodal transformer layer to derive the preliminary feature representation:
\begin{equation}
\begin{aligned}
(\hat{\mathcal{F}}_m\oplus \mathbf{\hat{v}}_{br})= Trans_v(\mathcal{F}_{m}\oplus \mathbf{v}_{br})
\end{aligned}
\end{equation}
where $\oplus$ stands for the concatenate operation. Then, the obtained bridge vector $\mathbf{\hat{v}}_{br}$ with spectral domain information is mapped to the feature space of the $\mathcal{F}_{t,a}$ through an MLP layer, and concatenated with the $\mathcal{F}_{t,a}$. The concatenated vectors are then fed into another Transformer layer to obtain the final fused features $\mathcal{F}_{m\to t,a}$:
\begin{equation}
\begin{aligned}
\mathcal{F}_{m\to t,a}= Trans_l(\mathcal{F}_{t,a} \oplus MLP(\mathbf{\hat{v}}_{br}))
\end{aligned}
\end{equation}

Symmetrically, using $\mathcal{F}_{t,a}$ as the initial input and applying the same operations, we can obtain the fused feature $\mathbf{\mathcal{F}}_{t,a\to m}$:
\begin{equation}
\begin{aligned}
(\hat{\mathcal{F}}_{t,a}\oplus \mathbf{\hat{v}'}_{br})= Trans_l(\mathcal{F}_{{t,a}}\oplus \mathbf{v}'_{br})
\end{aligned}
\end{equation}
\begin{equation}
\begin{aligned}
\mathcal{F}_{t,a\to m}= Trans_v(\mathcal{F}_{m} \oplus MLP(\mathbf{\hat{v}'}_{br}))
\end{aligned}
\end{equation}

Upon completing the multi-modal fusion, we employ $\mathcal{F}_{m\to t,a}$ and $\mathcal{F}_{t,a\to m}$ as inputs for two distinct linear classifiers and then average the logits for final classification.

\section{Experiments}
\subsection{Datasets and Baselines}
We conduct experiments on two datasets: IEMOCAP\cite{iemocap} and MELD\cite{meld}. IEMOCAP database is a widely used corpus in affective computing. It contains approximately 12 hours of audiovisual recordings and is designed for two-person dialogs. Each conversation in IEMOCAP has been segmented into utterances with the continuous label in the Valence-Arousal dimension and category label in categories such as $anger$, $happiness$, $sadness$, and $neutrality$. MELD is a popular dataset for tasks involving the analysis of emotions expressed by multiple speakers. It encompasses a collection of over 1400 dialogues and 13,000 speech instances extracted from the television show $Friends$. Emotion annotation in the dataset includes: $neutral$, $happiness$, $surprise$, $sadness$, $anger$, $disgust$, and $fear$. 

To showcase the efficacy of our proposed model across both unimodal and multi-modal scenarios, we selected the following baseline models for EPC task:
\begin{itemize}
\item \textbf{DEP}(2020)\cite{Shi20} incorporated the content from multiple speakers to enhance contextual information modeling.
\item \textbf{EAMT}(2023)\cite{Shi23} further considered multi-turn conversations and speaker information.
\end{itemize}
For ERC task, we selected several state-of-the-art models:
\begin{itemize}
\item \textbf{MM-DFN}(2022)\cite{mm-dfn} designed a new graph-based dynamic fusion module to fuse multi-modal context features.
\item \textbf{SCFA}(2023)\cite{scfa} proposed a novel speaker-aware cross-modal
fusion architecture.
\item \textbf{DF-ERC}(2023)\cite{df-erc} developed a disentanglement mechanism to decouple and model multi-modal features and the contextual aspects of conversations.
\end{itemize}

\subsection{Implementation} 
In this study, we utilize pre-trained wavLM\cite{wavlm} and RoBERTa models to extract 768-dimensional features from audio and text, respectively. The feature extraction module comprises two-layer Transformers, each with 8 heads and embeddings of 1024 dimensions. The Transformer in fusion-stage employs the same configuration. To ensure consistent output dimensions across modalities, we map the output of the prosody enhancement module to 1024 dimensions. Furthermore, we fine-tune the prosody encoder according to the configuration detailed in \cite{enc}. Within the TMF, we set the length of the $\mathbf{v}_{br}$ to 4 and stack two MFM blocks. The network is trained using the Adam optimizer, with a batch size of 32 and a learning rate of 0.0001.

\begin{table}[ht]
    \centering
	\caption{Comparison of performance characteristics on EPC task. $*$ indicates that our results are statistically significant under the t-test (p $<$ 0.05) relative to the comparison model. ``M'': Modality, ``T'': Text, ``S'': Speech}
        \begin{tabular}{cc|cc|cc}
            \hline
            \multicolumn{2}{c|}{Dataset}&\multicolumn{2}{c|}{IEMOCAP}&\multicolumn{2}{c}{MELD}\\
            \hline
            M & Method & UAR$\uparrow$ & M-F1$\uparrow$ & UAR$\uparrow$ & M-F1$\uparrow$\\
            \hline
            \multirow{3}*{T} 
            & DEP\cite{Shi20}& 74.96 & 74.54 & 42.19 & 42.67\\
            & EAMT\cite{Shi23} & 77.30 & 76.67 & 45.44 & 45.13 \\
            & Ours & \textbf{78.53*} & \textbf{78.06*} & \textbf{46.97*} & \textbf{46.23*} \\ 
            \hline
            \multirow{3}*{S} 
            & DEP\cite{Shi20}& 61.98 & 60.21 & 26.96 & 25.13\\
            & EAMT\cite{Shi23} & 65.01 & 65.91 & 28.67 & 25.97\\
            & Ours & \textbf{66.21*} & \textbf{66.32} & \textbf{29.12} & \textbf{27.85*}\\ 
            \hline
            \multirow{3}*{T$+$S} 
            & DEP\cite{Shi20}& 76.31 & 75.50 & 42.39 & 42.51\\
            & EAMT\cite{Shi23} & 80.18 & 80.01 & 45.21 & 44.36\\
            & Ours & \textbf{81.92*} & \textbf{81.53*} & \textbf{46.10} & \textbf{45.73*}\\ 
            \hline
        \end{tabular}
        \label{t1}
\end{table}

\begin{table}[ht]
    \centering
	\caption{Comparison of performance characteristics on ERC task. '-' means there is no relevant data in the source paper. The results with \underline{underline} denote suboptimal results.}
        \begin{tabular}{cc|cc|cc}
            \hline
            \multicolumn{2}{c|}{Dataset}&\multicolumn{2}{c|}{IEMOCAP}&\multicolumn{2}{c}{MELD}\\
            \hline
            M & Method & Acc$\uparrow$ & W-F1$\uparrow$ & Acc$\uparrow$ & W-F1$\uparrow$\\
            \hline
            \multirow{3}*{T} 
            & SCFA\cite{scfa} & 63.82 & 62.89 & 59.32 & 57.76\\
            & DF-ERC\cite{df-erc} & \textbf{65.13} & \textbf{65.46} & \textbf{65.17} & \textbf{64.54}\\
            & Ours & \underline{64.83} & \underline{64.32} & \underline{63.82} & \underline{63.01} \\ 
            \hline
            \multirow{3}*{S} 
            & SCFA\cite{scfa} & \textbf{49.24} & \textbf{48.66} & \textbf{47.37} & \textbf{44.18}\\
            & DF-ERC\cite{df-erc} & 41.47 & 38.62 & 43.83 & 41.72 \\
            & Ours & \underline{47.84} & \underline{47.06} & \underline{45.97} & \underline{42.94}\\ 
            \hline
            \multirow{3}*{T$+$S} 
            & MM-DFN\cite{mm-dfn} & - & 65.41 & - & 58.34 \\
            & SCFA\cite{scfa} & \textbf{67.91} & 66.42 & \textbf{64.86} & 63.69\\
            & Ours & \underline{67.79} & \textbf{66.62} & \underline{64.36} & \textbf{63.73} \\ 
            \hline
        \end{tabular}
        \label{t2}
\end{table}

\vspace{-5pt}

\begin{figure}[b]
	\centering
	\begin{subfigure}{0.45\linewidth}
		\centering
		\includegraphics[width=\linewidth]{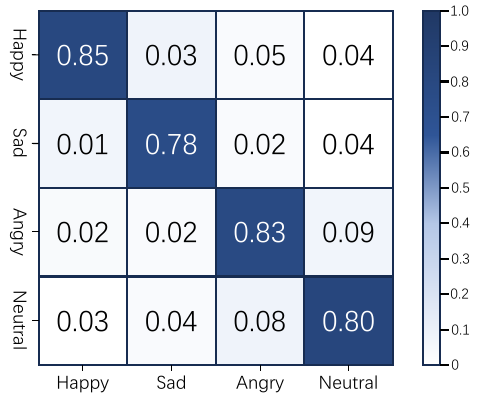}
		\caption{EPC}
        \label{f3a}		
	\end{subfigure}
	\begin{subfigure}{0.45\linewidth}
		\centering
		\includegraphics[width=\linewidth]{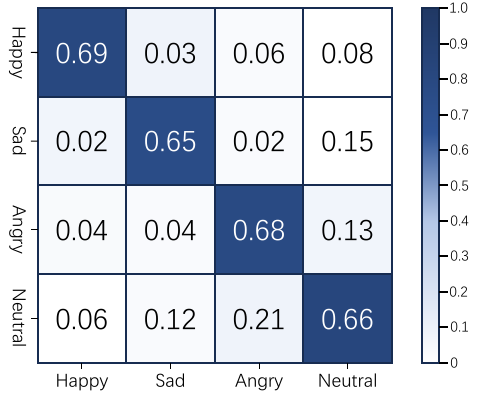}
		\caption{ERC}
        \label{f3b}
	\end{subfigure}
\caption{Accuracy confusion matrix on different tasks.}
\label{f3}
\end{figure}
\vspace{-25pt}

\subsection{Results analysis}
\textbf{Emotion Prediction} To compare with previous studies, we conducted experiments on the IEMOCAP and MELD datasets using the same evaluation metrics: Unweighted Average Recall (UAR) and Macro-F1 (M-F1), as employed in \cite{Shi20,Shi23}. The experimental results are depicted in Table~\ref{t1}. Our model outperforms previous studies across all three modal combinations and demonstrates statistical significance. In the text modality, excluding the two speech modal inputs, our model achieves UAR and M-F1 scores on IEMOCAP that surpass those of EAMT by 1.23\% and 1.39\%, respectively. This suggests that KWRT aids the model in perceiving emotional cues in text, thereby enhancing its capability for emotion prediction. Similarly, in the audio modality, our model also outperforms DEP's and EAMT's M-F1 scores on MELD by 2.72\% and 1.88\%. Additionally, to showcase the prediction results on specific emotion categories, we selected four emotions and constructed a confusion matrix based on IEMOCAP, as shown in Figure~\ref{f3a}. The results predicted by our model are mostly concentrated on the diagonal, indicating good performance in emotion prediction.\vspace{2pt}\\
\textbf{Emotion Recognition} To validate the model's performance on ERC tasks, we conducted experiments on the same datasets, using accuracy (Acc) and weighted F1 (W-F1) metrics, as shown in Table~\ref{t2}. Our model achieves comparable performance across all three modal combinations. Specifically, on MELD, our model outperforms SCFA's Acc by 4.50\% under the single text modality, although it is slightly lower than DF-ERC by 1.35\%. In the audio modality, our model performs comparably to the best-performing SCFA, surpassing DF-ERC by 2.14\%. In the multi-modal experiment, our model demonstrates performance comparable to MM-DFN and SCFA, indicating its strong emotion perception capabilities. Additionally, similar to the EPC task, we constructed a confusion matrix based on the IEMOCAP, as shown in Figure~\ref{f3b}. Our model continued to exhibit accurate performance.

\subsection{Ablation Study}
We conducted ablation experiments on the EPC task by removing the KWRT, PE, and TMF modules to demonstrate the effectiveness of our proposed components, as indicated in Table \ref{ab}. For example, on the IEMOCAP dataset, we first removed the KWRT module and directly used RoBERTa and the feature extraction module to output text features. UAR and M-F1 dropped by 0.93\% and 0.89\%, respectively, showing that the word-level annotation information in KWRT improves the emotional perception of text modalities. Next, we removed the PE module, resulting in UAR and W-F1 dropping by 0.75\% and 0.72\%. This indicates that although audio features contain emotional information, removing PE weakens the model's ability to perceive prosody. Finally, we removed the two-stage multi-modal fusion module based on the mel-spectrogram and retained only the preliminary fusion in the first stage, reducing UAR and M-F1 by 0.88\% and 0.84\%, respectively. This demonstrates that the new fusion method based on the mel-spectrogram enhances the model's ability to learn prosody information in the spectrum domain and improves the prediction effect.

\vspace{-5pt}
\begin{table}[h]
\centering
\caption{The results of ablation study on EPC task.}
\begin{tabular}{c|cc|cc}
    \hline
    Dataset&\multicolumn{2}{c|}{IEMOCAP}&\multicolumn{2}{c}{MELD}\\
    \hline
    & UAR$\uparrow$ & M-F1$\uparrow$ & UAR$\uparrow$ & M-F1$\uparrow$\\
    \hline
    Ours& /&/&/&/\\
    w/o KWRT& -0.93 & -0.89 & -1.12 & -1.10\\
    w/o PE & -0.75 & -0.72 & -0.71 & -0.69\\
    w/o TMF & -0.88 & -0.84 & -0.83 & -0.81 \\ 
    \hline
\end{tabular}
\label{ab}
\end{table}
\vspace{-5pt}

\section{Conclusion}
This article presents a two-stage multi-modal conversational emotion prediction model based on word-level relationship tagging, capable of perceiving contextual word-level information. At the same time, the designed prosody enhancement module captures prosodic information from the audio modality. In order to fuse spectral domain information, we employ the mel-spectrogram to perform two-step fusion process. Experiments have demonstrated the effectiveness of each proposed module and the superiority of the overall performance.

\section{Acknowledgement}
This work was supported by the Natural Science Foundation of China (62276242), National Aviation Science Foundation (2022Z071078001), CAAI-Huawei MindSpore Open Fund (CAAIXSJLJJ-2022-001A), Anhui Province Key Research and Development Program (202104a05020007), Dreams Foundation of Jianghuai Advance Technology Center (2023-ZM01Z001), USTC-IAT Application Sci. $\&$ Tech. Achievement Cultivation Program (JL06521001Y).

\bibliographystyle{IEEEtran}
\bibliography{ref}

\end{document}